\documentclass[preprint,nofootinbib,longbibliography,groupedaddress]{revtex4-2}
\usepackage{natbib}
\usepackage{hyperref}
\usepackage{amsmath} 
\usepackage{amssymb}
\usepackage{graphicx}
\usepackage{color} 
\usepackage{url}
\usepackage{tcolorbox}
\tcbuselibrary{fitting,raster,skins,xparse}
\usepackage{caption}
\usepackage{subcaption}

\usepackage{multirow}%

\begin{document}
	
\title{Unify the effect of anharmonicity in double-wells and anharmonic oscillators}

\author{Wei~Fan}
\author{Huipeng~Zhang}
\author{Zhuoran Li}
\email{zrl2lzr@gmail.com}
\affiliation{Department of Physics, School of Science, Jiangsu University of Science and Technology, Zhenjiang 212114, China}

\begin{abstract}
        We study the effect of anharmonicity in quantum anharmonic oscillators, by computing the energy gap between the ground  and the 1st excited state using the numerical bootstrap method. Based on perturbative formulae of limiting coupling regimes, we propose a qualitative formula of  the energy gap  across all coupling values. Except detailed numerical parameter values, the proposed formula  has the same functional form as the formula of ground state level splitting of double-well potentials, proposed recently in arXiv:2308.11516. This unifies the effects of anharmonicity in both the anharmonic oscillators and the double-well potentials, although the underlying physical process of them are completely different. We give an explanation of this connection of their anharmonicity from the viewpoint of quantum phase transitions. This connection is justified  up to the octic anharmonicities by the numerical bootstrap. 
\end{abstract}

\maketitle
\tableofcontents

\section{\label{sec:intro}Introduction}

In a recent paper~\cite{Fan:2023bld}, we studied the ground state level splitting $\Delta E_D$ caused by instantons in double-well potential models, which is an effect from the anharmonicity. We proposed a  qualitative formula for $\Delta E_D$ across all coupling values, so it describes both the 'weak' and the 'strong' regime of instanton effects. For the quartic case, it agrees with the known perturbative behaviors at limiting coupling regimes. The proposed formula is justified  on numerical data  for the quartic and the sextic double-well potential, obtained from the numerical bootstrap. Results suggest that the proposed formula works for  symmetric double-well potential 
of all anharmonicities, only with different numerical parameter values for different anharmonicities. 

In the present paper, we study the energy gap $\Delta E'$, due to anharmonicity, between the ground state and the 1st excited state of quantum anharmonic oscillators. This work is very natural, since the double-well potentials become anharmonic oscillators after changing the sign of coupling constant~\footnote{Here the coupling characterizes the strength of the anharmonicity effect: the stronger the coupling, the weaker the anharmonicity effect. This follows the convention of~\cite{Muller-Kirsten:2012wla}. 
} $g$ from minus to plus. Quantum anharmonic oscillators play an important role in perturbation theory~\cite{Muller-Kirsten:2012wla,Kleinert:788154}. In the weak regime of anharmonicity, the bare states are harmonic oscillators and the energy levels are asymptotic series, which is the famous large-order perturbation expansion firstly obtained by WKB methods~\cite{PhysRev.184.1231, PhysRevLett.27.461}. This asymptotic series is Borel summable~\cite{Graffi:1970erh} and is connected with the decay rate of inverted double-well potentials~\cite{Lipatov:1977hj,Brezin:1976vw,Brezin:1976wa,Zinn-Justin:1979ydh}.

It is an interesting question to ask \emph{what is the analytic structure of the strong anharmonicity regime}. In the strong regime, the bare states are unknown and the energy levels are only computed via numerical series. Here we propose a qualitative formula across all coupling values, connecting the two limiting regimes. The proposed formula here has \emph{the same functional form} as the one proposed in our recent paper~\cite{Fan:2023bld} for the ground state level splitting of double-well potentials, and this suggests a deep connection between the effects of anharmonicity in the anharmonic oscillators and in the double-well potentials. 

We test our proposed formula on numerical data up to the octic anharmonic oscillator. The data is computed by the non-perturbative numerical bootstrap method, where the  spectrum is solved from a set of consistency conditions that are built using only elementary properties of quantum theory without any perturbations.  It relies on the semidefinite programming algorithm~\cite{Poland:2011ey,Kos:2013tga,Kos:2014bka,Simmons-Duffin:2015qma} and  has rapidly developing applications in various fields of physics~\cite{Paulos:2016fap,Paulos:2016but,Paulos:2017fhb,Poland:2018epd,Poland:2022qrs,Anderson:2016rcw,Lin:2020mme,Han:2020bkb,Aikawa:2021qbl,Berenstein:2021dyf,Li:2022prn,Hu:2022keu,Nakayama:2022ahr,Nancarrow:2022wdr,Guo:2023gfi, John:2023him,Berenstein:2021loy,Bhattacharya:2021btd,Aikawa:2021eai,Tchoumakov:2021mnh,Bai:2022yfv,Khan:2022uyz,Berenstein:2022ygg,Morita:2022zuy,Blacker:2022szo,Berenstein:2022unr,Berenstein:2023ppj,2020arXiv200606002H,Lawrence:2021msm, Hessam:2021byc,Kazakov:2021lel,Kazakov:2022xuh,Du_2022,Lin:2023owt}. For applying the numerical bootstrap to anharmonic oscillators, the algorithm has been studied in great detail by ~\cite{Han:2020bkb,Aikawa:2021qbl,Berenstein:2021dyf,Li:2022prn,Hu:2022keu,Nakayama:2022ahr,Nancarrow:2022wdr, Guo:2023gfi, John:2023him}. As in~\cite{Fan:2023bld}, here we follow the implementation introduced firstly by~\cite{Aikawa:2021qbl}.

The paper is organized as follows. In Section~\ref{sec:bootstrap}, we briefly review steps of the numerical bootstrap method. In Section~\ref{sec:oscillator} we define the models and analyze the  bootstrap data of anharmonic oscillators. We propose the qualitative formula~\eqref{eq:formulaP} after introducing the one-loop formula~\eqref{eq:H4E0} of the quartic anharmonic oscillator, making it convenient to discuss the detailed physics there. In Section~\ref{sec:unify}, we explain the connection with the double-well potentials. We analyze the effects of anharmonicity in the anharmonic oscillators and the double-well potentials from the view point of quantum phase transitions. In Section~\ref{sec:conclusion}, we close  with a discussion of open questions and future work.

\section{\label{sec:bootstrap}Method of the numerical bootstrap}

For the reason of completeness, here we repeat the brief review of numerical bootstrap already given in our previous paper~\cite{Fan:2023bld}. The readers familiar with it may skip this part.  For the practical implementation, we  follow the choice of operators $\{x^mp^n\}$ introduced firstly by~\cite{Aikawa:2021qbl}. The algorithmic details have been analyzed extensively from various aspects, and the interested reader can go to references listed in the Introduction. 

The numerical bootstrap divides into three  steps: (1) choose a set of operators and derive their recursive equations, (2)  impose positive constraints (or other working constraints) on the operators and obtain the bootstrap matrix, (3) set the numerical search space and use the semidefinite optimization to find allowed parameter values in this space. 
Here for anharmonic oscillators, we choose the system energy as the search parameter, because our aim is to compute the energy of the ground state and the first excited state.

\subsection{Recursive equations}

   For a quantum mechanical system, we use the following Hamiltonian:
\begin{equation}\label{one}
        H=\frac{p^2}{2}+V(x)
\end{equation}
For its energy eigenstate  $\left| \varphi  \right\rangle$ with eigenvalue $E$, the expectation value of operators composed of $H$ and an arbitrary operator $\alpha $, must satisfy the following two constrains, 
\begin{equation}\label{two}
       \left\langle {\left[ {H,\alpha } \right]} \right\rangle \equiv \left\langle  \varphi | {\left[ {H,\alpha } \right]} | \varphi \right\rangle =   0, \,
       \left\langle {H\alpha } \right\rangle  = \left\langle {\alpha H} \right\rangle  = E\left\langle \alpha  \right\rangle 
\end{equation} 
If choose $\alpha  = {x^n}, {x^n}p$, these constrains lead to the following recursion relations~\cite{Aikawa:2021qbl,Hu:2022keu}
\begin{align}\label{sev}
 n(n - 1)(n - 2)&\left\langle {{x^{n - 3}}} \right\rangle  - 8n\left\langle {{x^{n - 1}}V(x)} \right\rangle 
 + 8nE\left\langle {{x^{n - 1}}} \right\rangle  - 4\left\langle {{x^n}{V'}(x)} \right\rangle  = 0.   
\end{align}

\subsection{bootstrap matrix}
To build the bootstrap matrix, we follow the efficient construction proposed by~\cite{Aikawa:2021qbl} by choosing the following operators,
\begin{equation}\label{ten}     
 \mathcal{O}  = \sum\limits_{m = 0}^{k_x}\sum\limits_{n = 0}^{k_p} {{C_{mn}}{x^m}{p^n}}
\end{equation}
with ${C_{mn}}$ being constants. The expectation value of the operator $\mathcal{O}$ on the energy eigenstate $\left| \varphi  \right\rangle $  satisfies the positivity constraint
\begin{equation}\label{ele}
         \left\langle {{{\mathcal{O}} ^\dag }{\mathcal{O}} } \right\rangle \geq 0.
\end{equation}
From this constraint, the bootstrap matrix can be built that must be positive-semidefinite
\begin{equation}\label{eq:shisan}
        {M}: = \left( {\begin{array}{*{20}{c}}
{{\mathcal{O}}_0^{\dag }{{\mathcal{O}}_0}} & {{\mathcal{O}} _0^{\dag }{{\mathcal{O}} _1}}& \ldots &{{\mathcal{O}} _0^{\dag }{{\mathcal{O}} _k}}\\
{{\mathcal{O}} _1^{\dag }{{\mathcal{O}} _0}}&{{\mathcal{O}} _1^{\dag }{{\mathcal{O}} _1}}& \ldots &{{\mathcal{O}} _1^{\dag }{{\mathcal{O}} _k}}\\
 \vdots & \vdots & \ddots & \vdots \\
{{\mathcal{O}} _k^{\dag }{{\mathcal{O}} _0}}&{{\mathcal{O}} _k^{\dag }{{\mathcal{O}} _1}}& \ldots &{{\mathcal{O}} _k^{\dag }{{\mathcal{O}} _k}}
\end{array}} \right)  \geq 0,
\end{equation}
where $\mathcal{O}_{0,1,2,\ldots}$ are the component operators of $\mathcal{O}$ and the maximum value $k$ is called the depth of the bootstrap matrix. As the depth $k$ increases, the positive-semidefinite constraint becomes stronger and so the numerical results becomes more accurate.

\subsection{Search space}

The last step is to set the search space  which is a minimum set of data to initialize the recursion. After choosing search parameters and the optimization target, the positive-semidefinite optimization will exclude  parameter values that do not satisfy the constraints~\eqref{eq:shisan} and so reduce the size of the parameter space. With sufficiently large depth $k$, the remaining parameter space passing the constraints~\eqref{eq:shisan} will be a tiny neighborhood that can be viewed as a data point. This data point is the discrete eigenvalue of the quantum system with the optimized target value being the expectation. The search space of the quartic anharmonic oscillator is $\left\{ E,{\rm{ }}\left\langle {{x^2}} \right\rangle  \right\}$, the sextic case is $\left\{ {E,{\rm{ }}\left\langle {{x^2}} \right\rangle ,{\rm{ }}\left\langle {{x^4}} \right\rangle } \right\}$ and the octic case is $\left\{ {E,{\rm{ }}\left\langle {{x^2}} \right\rangle ,{\rm{ }}\left\langle {{x^4}} \right\rangle, \left\langle {{x^6}} \right\rangle } \right\}$. 

\begin{figure}[b]
\centering
\includegraphics[width=1\linewidth]{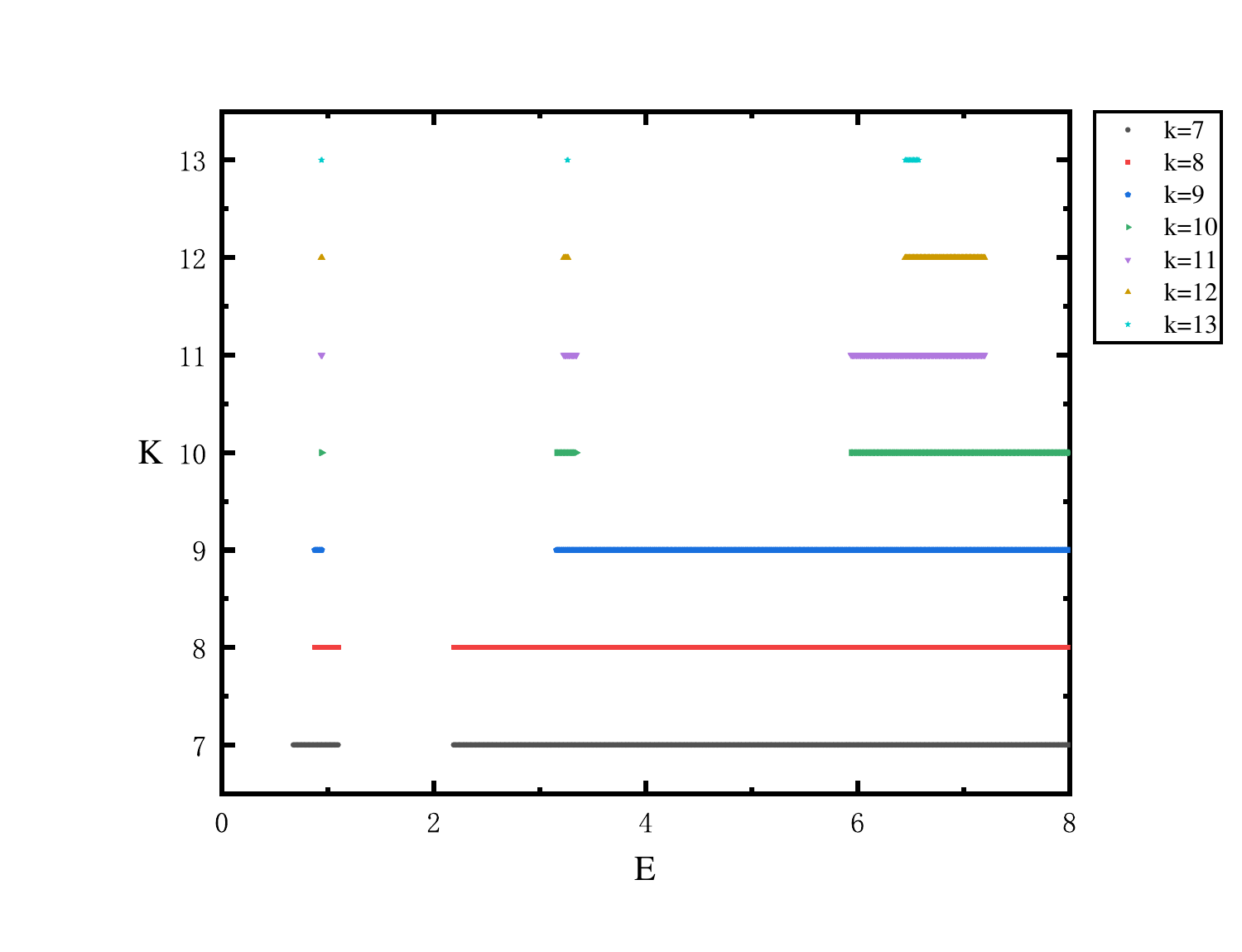}
\caption{An example bootstrap data of the octic oscillator at $g=1$ for various depth $k$ of bootstrap matrix. As the depth $k$ increases from $7$ to $13$, the first two eigenvalues $E_0$ and $E_1$ become more and more accurate.  }\label{fig:band8} 
\end{figure}

An example of the bootstrap data is shown in Figure~\ref{fig:band8} for the octic oscillator at $g= 1$. For small depth $k=7$, the ground state is separated from excited states, but the resolution  is very low. As the depth increases to $k=13$, the ground state  and the first excited states are identified to a high resolution, and higher excited states are separated out. At this depth $k=13$ the bootstrap matrix is not large, so this implementation~\cite{Aikawa:2021qbl} is very efficient. This behavior with depth $k$ is the same as that in~\cite{Fan:2023bld} for double-well potentials. 

\section{\label{sec:oscillator}Bootstraping anharmonic oscillators}

The pure anharmonic oscillator is defined as 
\begin{equation}
\label{eq:Hn}
H=\frac{p^2}{2} + g x^2 + x^{2n}, \quad g > 0 \,\& \, n\geq 2.
\end{equation}
When $g<0$, this becomes the double-well potential studied in~\cite{Fan:2023bld}. 
In this convention, the coupling $g$ is of the 'mass' term~\footnote{This follows the convention of~\cite{Muller-Kirsten:2012wla} where the asymptotic series  is expanded in terms of the large mass term $h^2 x^2$.}: large coupling $g\gg 1$ is the weak regime of anharmonicity   and small coupling $g\to 0$ is the strong regime of anharmonicity.
From the energy gap $\Delta E= E_1 -E_0$ between the ground  and the 1st excited state, we extract the contribution of anharmonicity by
\begin{align}
\label{eq:gapDef}
\Delta E' = \Delta E - \sqrt{2 g},
\end{align}
where  $\sqrt{2g}$ is the energy gap of harmonic oscillators~\footnote{In our convention the frequency of the harmonic oscillator is $\hbar \omega=\sqrt{2g}$}. It is a standard practice to separate out the contribution of harmonic oscillators, when studying the energy levels of anharmonic oscillators~\cite{Muller-Kirsten:2012wla,Kleinert:788154}.

In the weak regime of the quartic anharmonic oscillator, the asymptotic series of $E_0$ can be recast as a closed formula by the one-loop path integral~\cite{Kleinert:788154} 
\begin{equation}
\label{eq:H4E0}
E_0(g)= \sqrt{\frac{g}{2}}+\frac{\sqrt{6} g-\sqrt{2 \pi } e^{g^3/3} g^{5/2} \operatorname {erfc}\left(\frac{g^{3/2}}{\sqrt{3}}\right)}{\pi },
\end{equation}
where $\operatorname {erfc}(z)$ is the complementary error function. In the harmonic limit $g\to \infty$, the asymptotic expansion of $\operatorname {erfc}(z)$ ~\cite{NIST:DLMF}
\begin{equation}
  \operatorname {erfc}(z)  \displaystyle\sim\frac{e^{-z^{2}}}{\sqrt{\pi}}\sum_{m=0}^{\infty}(-1)^{m}\frac{{\left(\tfrac{1}{2}\right)_{m}}}{z^{2m+1}}
\end{equation}
gives the correct ground state energy of harmonic oscillators
\begin{equation}
E_0(g)\displaystyle\sim \sqrt{\frac{g}{2}}+\frac{3 \sqrt{\frac{3}{2}}}{\pi  g^2}+O\left(\frac{1}{g^{9/2}}\right) \sim \sqrt{\frac{g}{2}}.
\end{equation} 
Higher energy levels can be computed as a  numerical series  using the variational approach to tunneling~\cite{Kleinert:1992tq}, but there is no closed formula like~\eqref{eq:H4E0}. In the strong regime of anharmonicity, $g\to 0$ of ~\eqref{eq:H4E0} gives $E_0=0$ which is obviously wrong. So the one-loop perturbation~\eqref{eq:H4E0} can not describe the strong regime. In the strong regime, energy levels are usually given as numerical convergent series. 

From an educated guess, we propose the following formula of the energy gap $\Delta E'$ due to the anharmonicity
\begin{equation}
    \label{eq:formulaP}
    \Delta E'(g) = 2 \Delta E'(0) \frac{e^{-a g^{b}}}{1+e^{- c g^d}}, \quad   a, b, c, d > 0,
\end{equation}
with numerical parameter values depending on the anharmonicity order $n$. 
In the harmonic limit $g\to\infty$, this anharmonicity approaches zero
\begin{align}
  \Delta E'(g) \sim e^{-a g^{b}} \to 0,\quad g\to\infty,
\end{align}
which is expected for the hamonic oscillator. 
In the strong regime $g\to 0$ of anharmonicity, this is a convergent series of $g$
\begin{align}
 \Delta E'(g) \sim  \Delta E'(0)\left(1 - a g^b + \frac{c}{2} g^d +\ldots \right),\quad g\to 0,
\end{align}
which is  expected because this limit is usually studied by numerical series~\cite{Muller-Kirsten:2012wla,Kleinert:788154} of $g$. 
Furthermore, the propopsed formula happens to be the  formula proposed in~\cite{Fan:2023bld} for the ground state level splitting of double-well potentials. This suggests that the effect of anharmonicity between them is connected, which will be discussed in Section~\ref{sec:unify}.

\subsection{Quartic oscillator}
\label{sec:quartic}
\begin{figure}
        \centering
\begin{subfigure}[b]{0.46\textwidth}
  \centering
\includegraphics[width=\linewidth]{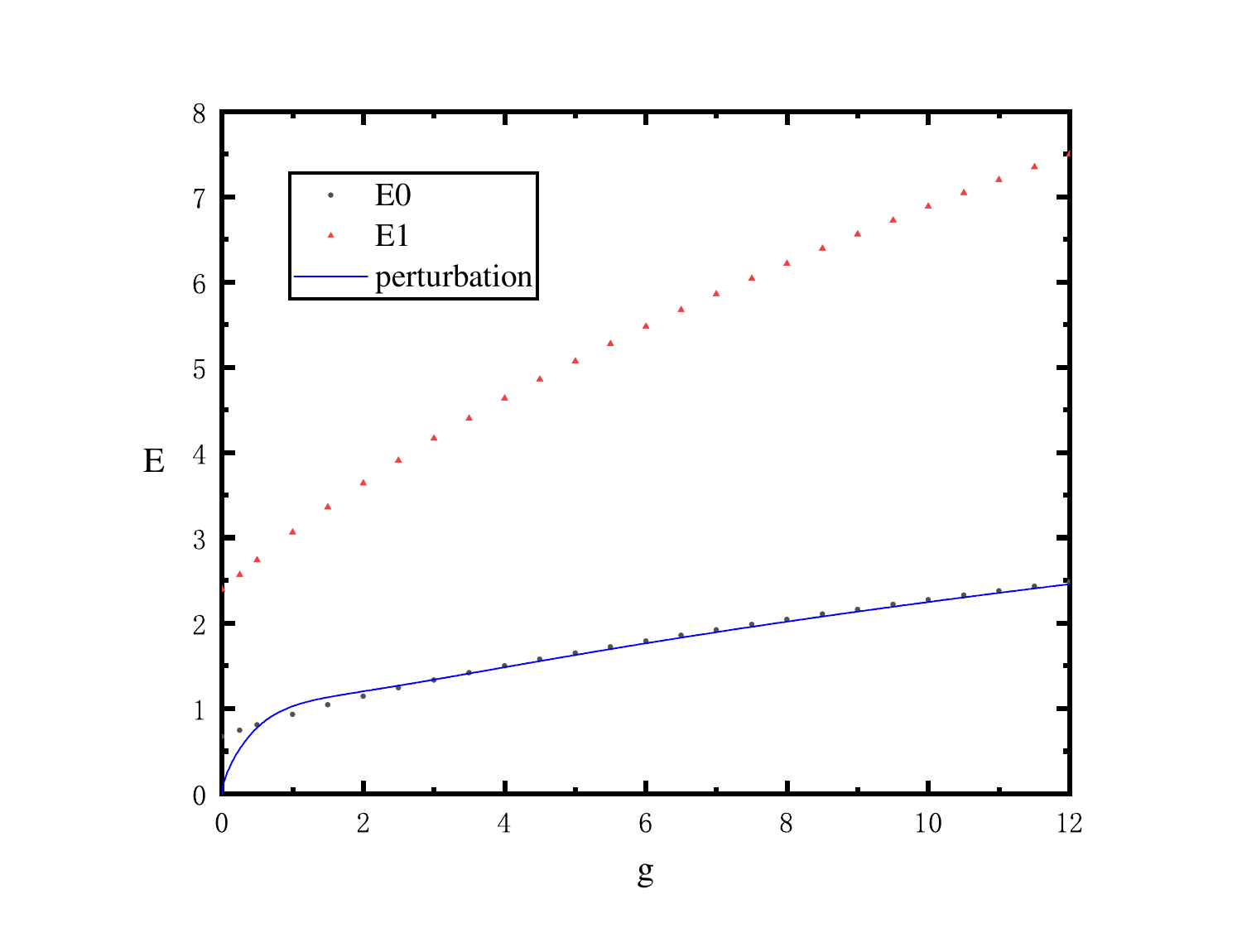}
  \caption{$E_0$ and $E_1$ of the quartic oscillator. The solid line is the perturbation~\eqref{eq:H4E0}.}
 \label{fig:E4}
\end{subfigure}
   \hfill
\begin{subfigure}[b]{0.46\textwidth}
\centering
\includegraphics[width=\linewidth]{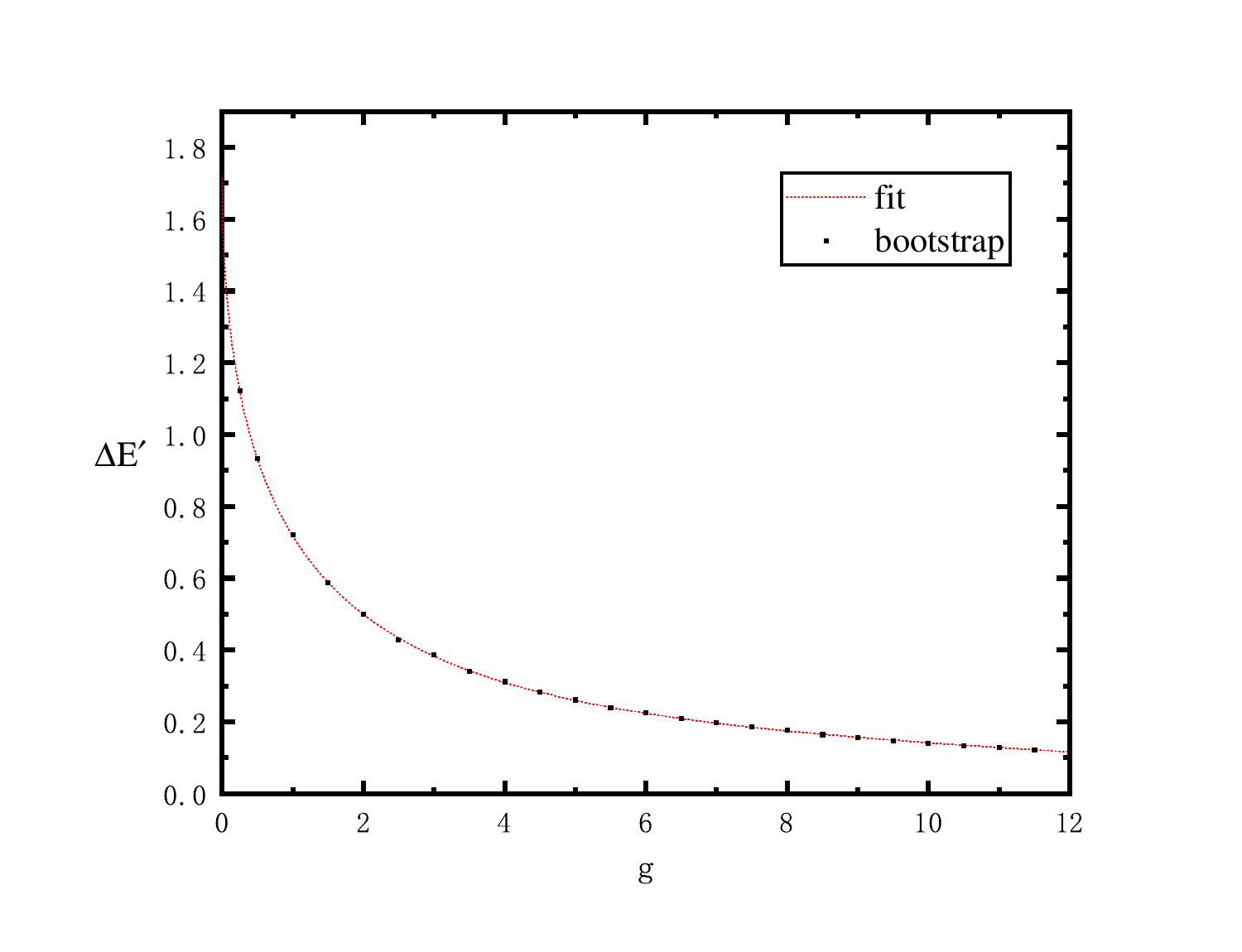}
\caption{$\Delta E'$ of the quartic oscillator. The solid line is the propopsed formula~\eqref{eq:formulaP}.}
\label{fig:fit4}
\end{subfigure}
\caption{Results for the quartic oscillator. In the harmonic limit of large $g$, the perturbation calculation captures the correct physics. But the perturbation fails at the strong anharmonicity regime $g\to 0$. The propopsed formula ~\eqref{eq:formulaP} correctly describes the energy gap $\Delta E'$. }
\label{fig:quartic}
\end{figure}
For the quartic oscillator, Figure~\ref{fig:E4} shows the bootstrap data of $E_0$ and $E_1$, and the  perturbation formula~\eqref{eq:H4E0} of $E_0$. In the harmonic limit $g\gg 1$, the perturbation formula approaches the bootstrap data with a tiny deviation, so perturbation captures the majority, but not all, of the physics of the ground state. In the strong regime $g\to 0$, the perturbation formula of asymptotic series~\eqref{eq:H4E0} fails completely. This strong regime can be described by convergent numerical series~\cite{Muller-Kirsten:2012wla,Kleinert:788154} of $g$, but the convergent series fails in the weak regime. This is expected of perturbation theory that it is valid only  near the regime where it is expanded. 

Figure~\ref{fig:fit4} shows the energy gap $\Delta E'$ due to the anharmonicity and the proposed formula $\Delta E'(g)$~\eqref{eq:formulaP} fitted to the data. The data $\Delta E'$ is maximum at $g=0$ and then quickly drops. When $g\gg 1$, it goes to zero asymptotically. This asymptotic behavior is the reason of the tiny deviation between the data $E_0$ and the perturbative formula~\eqref{eq:H4E0} in the harmonic limit, shown in Figure~\ref{fig:E4}. This asymptotic decreasing behavior of $\Delta E'$ is expected and it explains why perturbative energy levels in the weak regime are asymptotic series like~\eqref{eq:H4E0}. The proposed formula $\Delta E'(g)$~\eqref{eq:formulaP} fits well with the data, with parameter values given in Table~\ref{table:fit}. This justifies  the qualitative formula~\eqref{eq:formulaP}.

\subsection{Sextic oscillator}
\label{sec:sextic}
\begin{figure}
        \centering
        \begin{subfigure}[b]{0.46\textwidth}
                \centering
                \includegraphics[width=\linewidth]{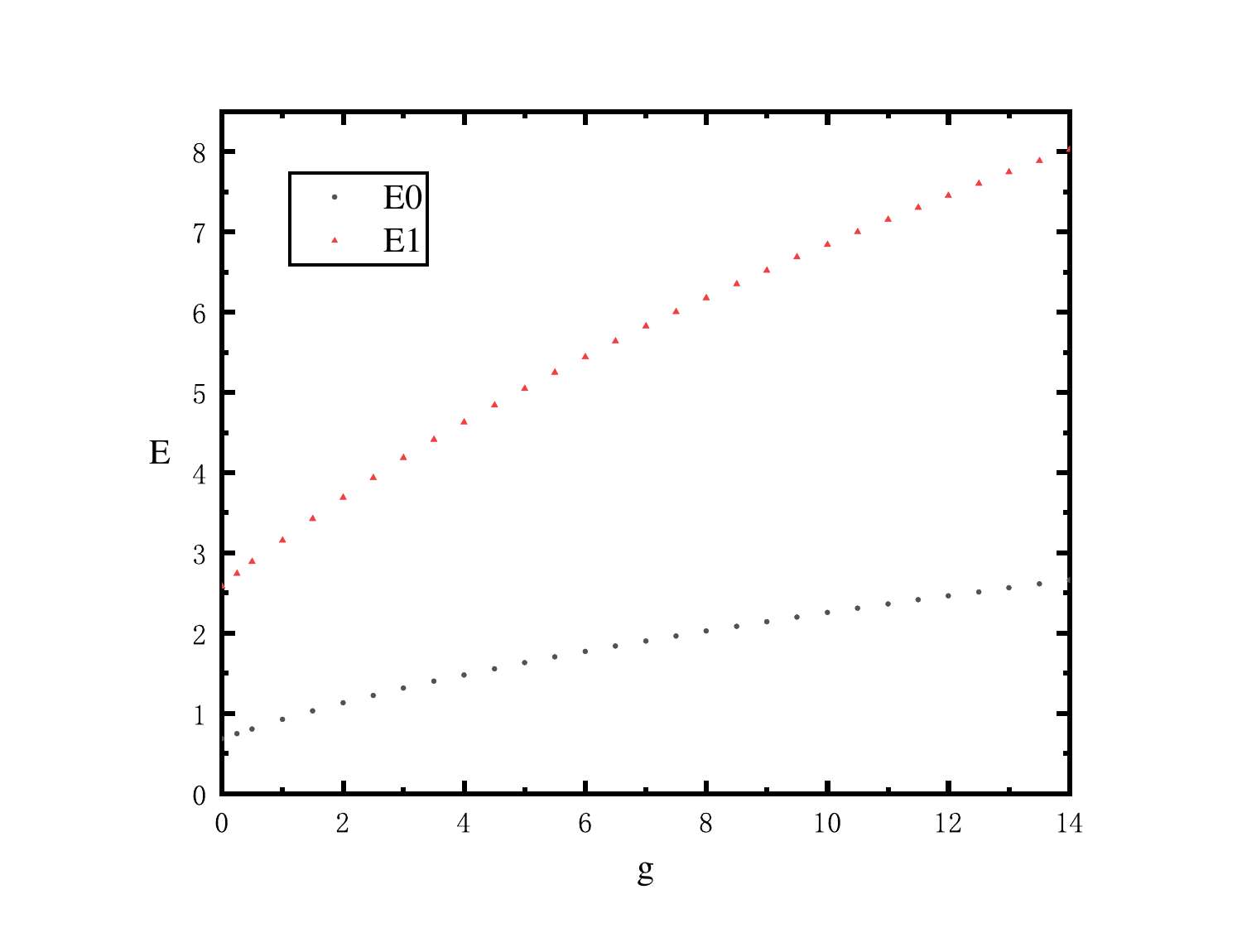}
                \caption{$E_0$ and $E_1$ of the sextic oscillator.}
                \label{fig:E6}
        \end{subfigure}
           \hfill
        \begin{subfigure}[b]{0.46\textwidth}
                \centering
                \includegraphics[width=\linewidth]{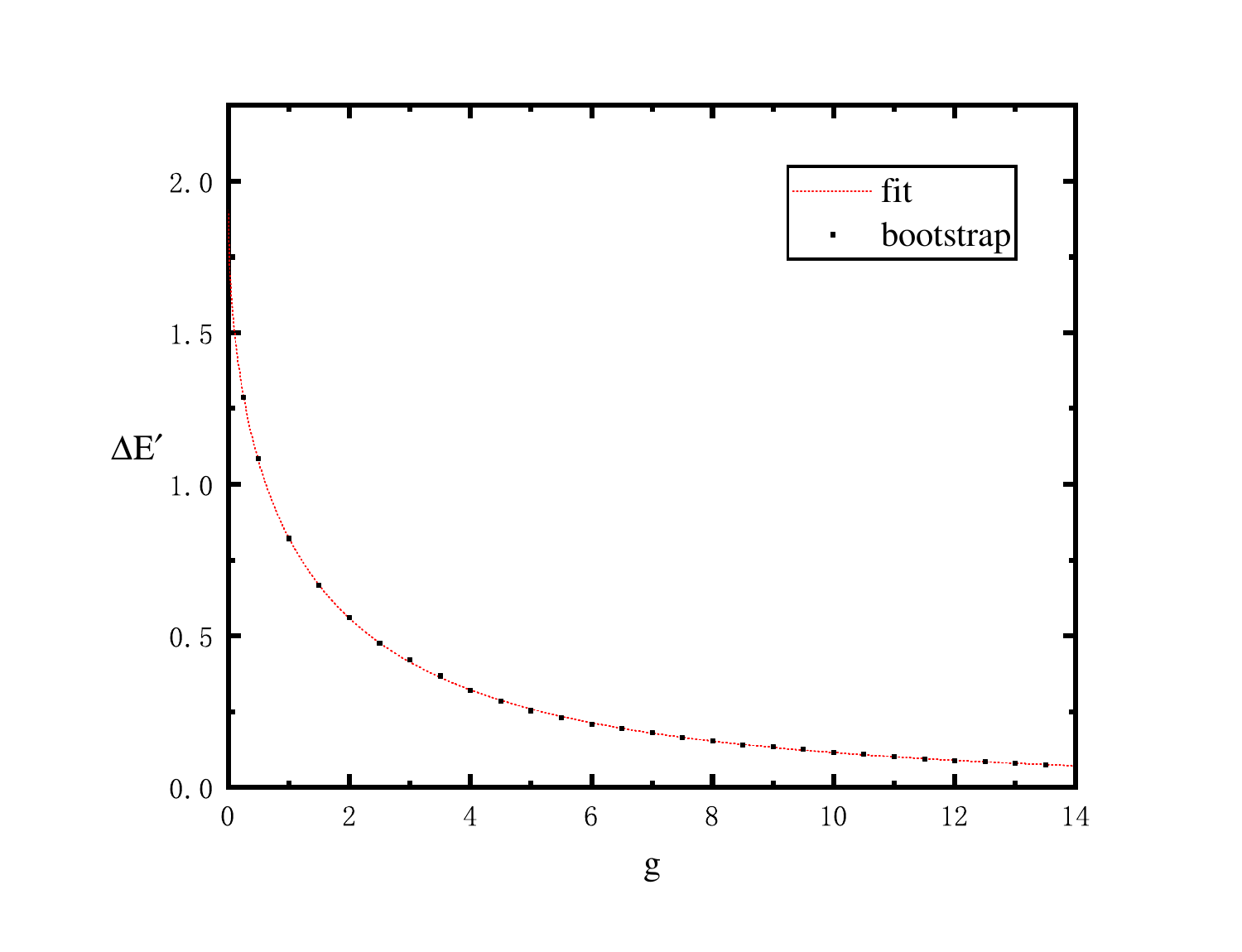}
                 \caption{$\Delta E'$ of the sextic oscillator.}
                 \label{fig:fit6}
        \end{subfigure}
\caption{Results for the sextic oscillator. The propopsed formula ~\eqref{eq:formulaP} correctly describes the energy gap $\Delta E'$ across the small and the large coupling regimes.}
\label{fig:sextic}
\end{figure}
For the sextic oscillator, Figure~\ref{fig:E6} shows the bootstrap data of $E_0$ and $E_1$, and Figure~\ref{fig:fit6} shows the energy gap $\Delta E'$ due to the anharmonicity. We see similar behavior of $\Delta E'$ as in the quartic case. The proposed formula $\Delta E'(g)$~\eqref{eq:formulaP} fits well with the data, with parameter values given in Table~\ref{table:fit}. 

\subsection{Octic oscillator}
\label{sec:octic}
\begin{figure}
        \centering
        \begin{subfigure}[b]{0.46\textwidth}
                \centering
                \includegraphics[width=\linewidth]{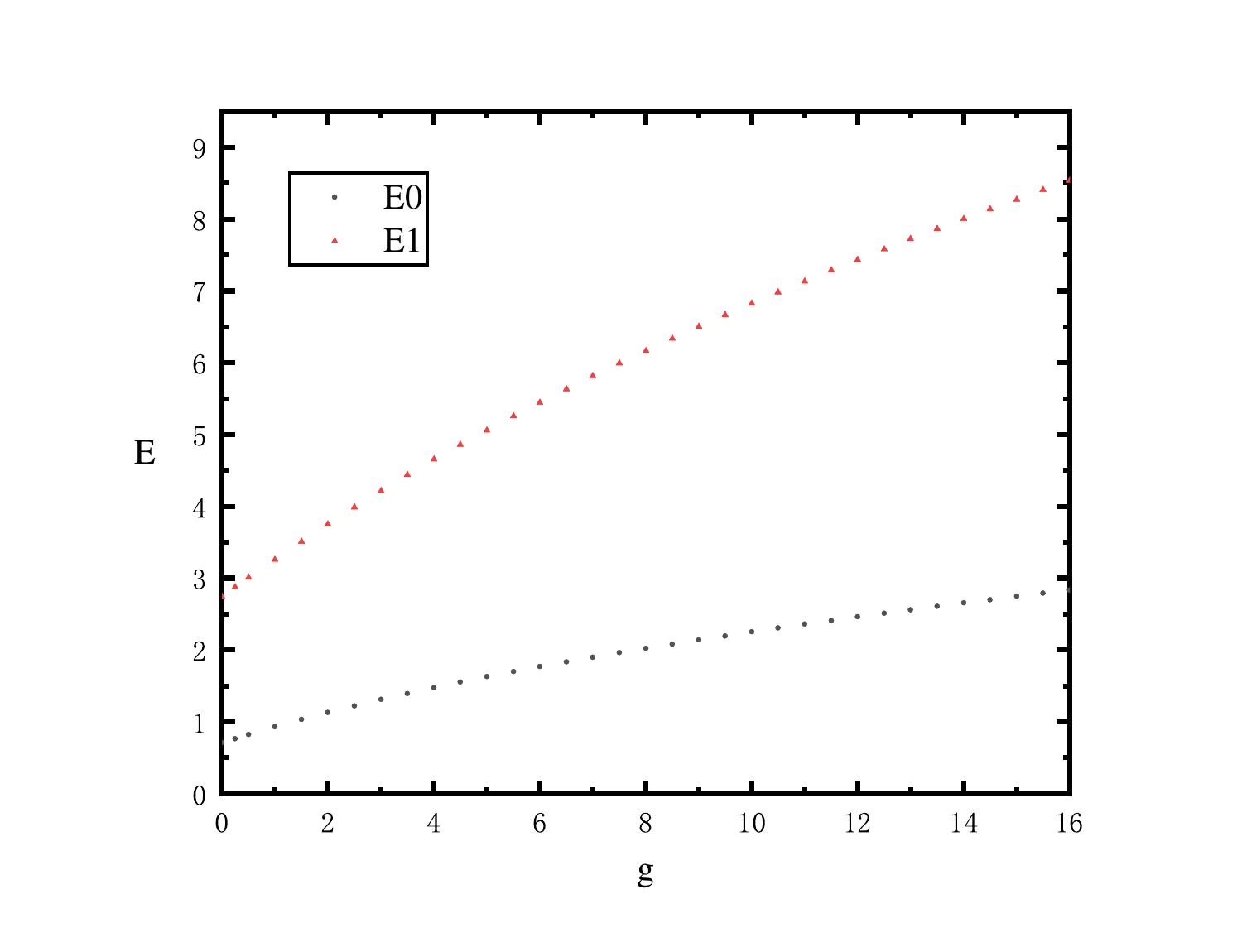}
                \caption{$E_0$ and $E_1$ of the octic oscillator.}
                \label{fig:E8}
        \end{subfigure}
           \hfill
        \begin{subfigure}[b]{0.46\textwidth}
                \centering
                \includegraphics[width=\linewidth]{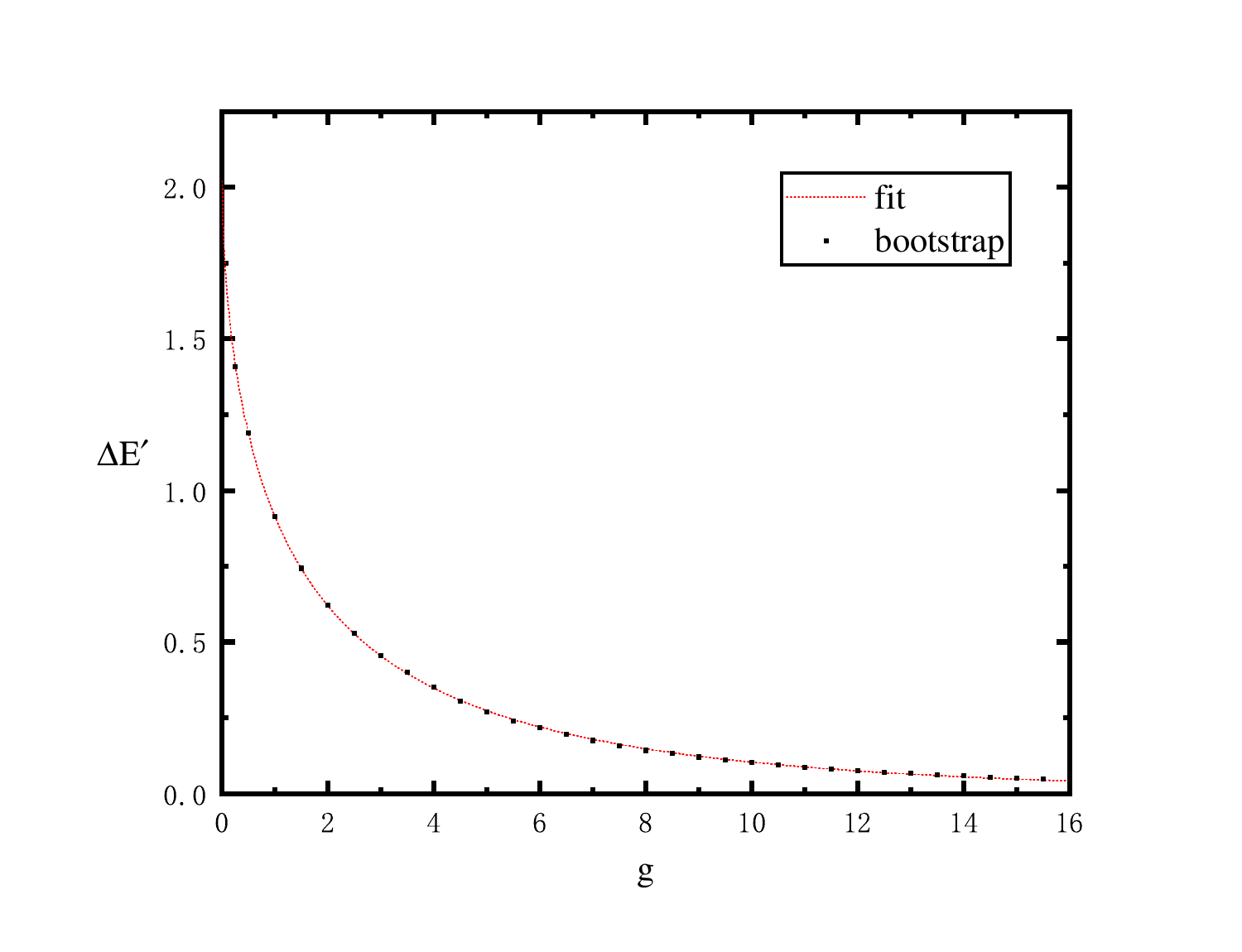}
                 \caption{$\Delta E'$ of the octic oscillator.}
                 \label{fig:fit8}
        \end{subfigure}
        \caption{Results for the octic oscillator.  The propopsed formula ~\eqref{eq:formulaP} correctly describes the energy gap $\Delta E'$ across the small and the large coupling regimes.}
        \label{fig:octic}
\end{figure}

For the octic oscillator, Figure~\ref{fig:E8} shows the bootstrap data of $E_0$ and $E_1$, and Figure~\ref{fig:fit8} shows the energy gap $\Delta E'$ due to the anharmonicity. We see similar behavior of $\Delta E'$ as in the quartic case. The proposed formula $\Delta E'(g)$~\eqref{eq:formulaP} fits well with the data, with parameter values given in Table~\ref{table:fit}. 

For the anharmonic oscillators, the fitted parameter values are in Table~\ref{table:fit}. We see that the parameter $b$ is increasing with anharmonicity order $n$ and the parameter $a,c,d$ are descreasing with the anharmonicity order $n$.  These results suggest that the parameter values are not random and there is a connection with the anharmonicity order $n$. This unknown connection might be related with a 'critical behavior' which will be explained in the next section.

\begingroup
\begin{table}[ht]
\centering
\begin{tabular}{|c | c   c |} 
 \hline
\multirow{2}{*}{n}  & a & b \\
&c & d  \\ 
 \hline
\multirow{2}{*}{2} & 0.8842629827697163 &0.5193664589411509\\
  & 0.015270747582868331 &  1.888568625833652 \\
          \hline
\multirow{2}{*}{3}  & 0.8421595166826806 &0.5597339573586666\\
 & 0.008140628665542498 & 1.8560295155792268\\
          \hline
\multirow{2}{*}{4} &  0.8014327779076852&0.5720036705632807\\
 &0.0015383208190051754 & 1.3556010103295226 \\
         \hline
        \end{tabular}
        \caption{Fitted parameter values of~\eqref{eq:formulaP} for anharmonic oscillators with anharmonicity $2n$.}
        \label{table:fit}
\end{table}
\endgroup

\section{\label{sec:unify}Unifying with double-well potentials}

\begin{figure}
\centering
\includegraphics[width=\linewidth]{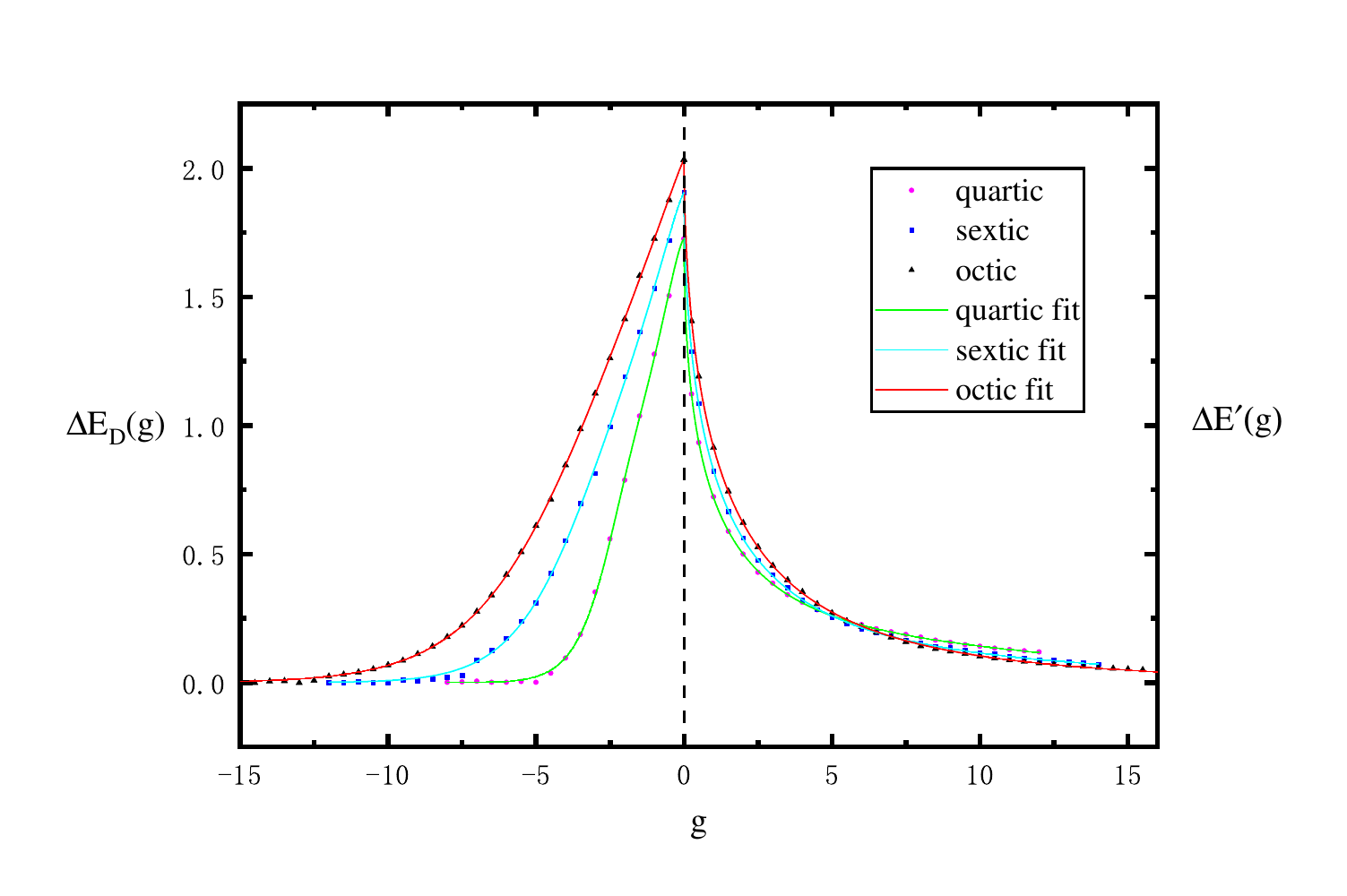}
\caption{$\Delta E'(g)$ and $\Delta E_D(g)$ for the quantum system $V(x)= g x^2 + x^{2n}$.  When viewed as the order parameter, they describe the two different 'phases' with $g>0$ and $g<0$ respectively, where the transition point is at $g=0$. They are continuous at $g=0$, but their derivatives are not, which is shown in Figure~\ref{fig:susceptibility}.}\label{fig:all}
\end{figure}
\begin{figure}
        \centering
        \includegraphics[width=\linewidth]{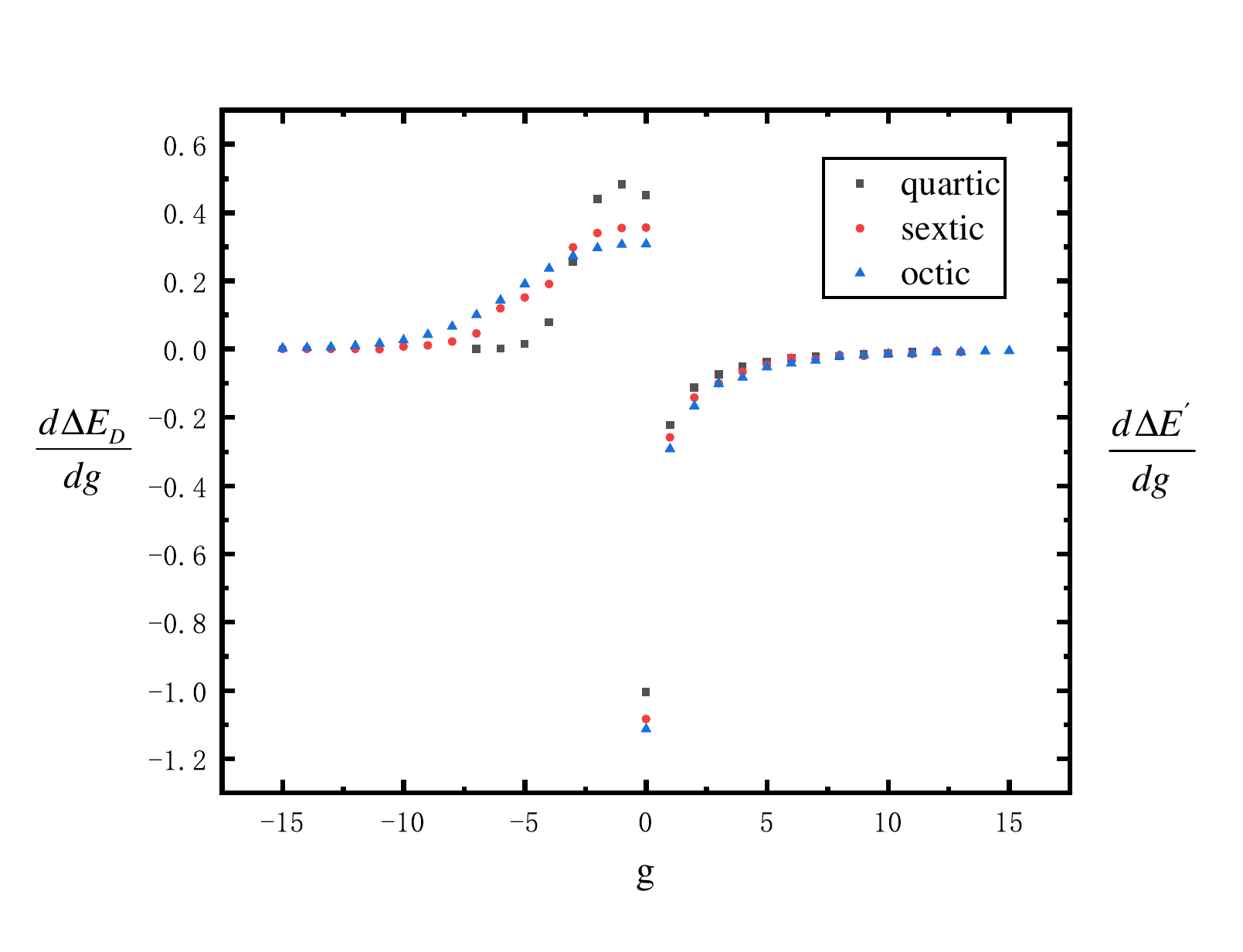}
        \caption{$\frac{d \Delta E'}{dg}$ and $\frac{d \Delta E_D}{d g}$ for the quantum system $V(x)= g x^2 + x^{2n}$. These first-order derivatives can be viewed as an analog of 'susceptibility'. As $n$ increases, it tends to diverge from the right hand side when approaching the transition point $g=0$.}\label{fig:susceptibility}
 \end{figure}
The proposed formula $\Delta E'(g>0)$~\eqref{eq:formulaP} here for anharmonic oscillators  is the same  formula $\Delta E_D(g<0)$ proposed in~\cite{Fan:2023bld} for double-well potentials. To understand this unexpected connection, let's treat anharmonic oscillators and double-well potentials as a single quantum system $V(x)= g x^2 + x^{2n}$, then $g>0$ and $g<0$ can be viewed as two phases with oscillator physics and instanton physics respectively. 

Like in quantum phase transitions, the energy gap due to anharmonicity can be viewed as the order parameter~\cite{Sachdev_Subir2011-05-09} that characterizes the behavior of the system.
\begin{itemize}
\item  For anharmonic oscillators $g>0$, $\Delta E'$ is the contribution of anharmonicity, because the contribution $\sqrt{2g}$ of hamonic oscillators has been subtracted by definition~\eqref{eq:gapDef}. In the limiting weak regime of anharmonicity $g\to\infty$, $\Delta E'\to 0$, so $\Delta E'(g)$ captures the deviation from this limiting regime and is an effect of anharmonicity.  
\item For  double-well potentials $g<0$, the energy gap  $\Delta E$ at finite $g$ is also called the ground state level splitting due to instantons. In the limiting weak regime of anharmonicity, the double-well potential has zero level splitting, so we can treat $\Delta E_D(g)=\Delta E(g) - 0$ as capturing the deviation from this limiting regime and is an effect of anharmonicity, the same as $\Delta E'(g)$ of anharmonic oscillators.  
\end{itemize}

With this view of $\Delta E'(g)$ and $\Delta E_D(g)$, the proposed formula~\eqref{eq:formulaP} characterizes the effect of anharmonicity in this system. So  $\Delta E'(g)$ and $\Delta E_D(g)$ can be viewed as the order parameter in the two phases of the system.  Figure~\ref{fig:all} shows them together with data  up to the octic case.  The data of double-well potentials is taken from~\cite{Fan:2023bld}. In the phase $g<0$, it shows the data of $\Delta E_D(g)$ coming from instanton physics.  In the phase $g>0$, it shows the data of $\Delta E'(g)$ coming form oscillator physics. In both phases, the proposed formula~\eqref{eq:formulaP} agrees well with the data, which suggests a deep connection between anharmonic oscillators and double-well potentials. 
Furthermore, they all look like a $\lambda\operatorname{-shape}$ curve that is common in phase transitions~\cite{BUCKINGHAM196180}. So this implies a strong analog with phase transitions. With this viewpoint, the parameter values in Table~\ref{table:fit} might be related with 'critical exponents'. 

If $\Delta E'(g)$ and $\Delta E_D(g)$ are viewed as the order parameter in the two phases of the system, their derivatives can be viewed as the susceptibility of the order parameter. Figure~\ref{fig:susceptibility} shows this 'susceptibility'. We see that as $n$ increases, the derivative diverges from the right hand side when approaching $g=0$, which is the transition point between the two 'phases' of the system. This divergence of 'susceptibility' at the transition point suggests again a possibility of quantum phase transition.

\section{\label{sec:conclusion}Conclusion}

In this work, we studied the energy gap $\Delta E'(g)=E_1-E_0 - \sqrt{2g}$ originated from anharmonicity of quantum anharmonic oscillators, where the contribution $\sqrt{2g}$ of pure harmonic oscillators is subtracted off. We used the numerical bootstrap method following~\cite{Aikawa:2021qbl} for the detailed implementation of the algorithm. We proposed a qualitative formula~\eqref{eq:formulaP} for $\Delta E'(g)$ across all coupling values, with the expected behavior both in the strong  and the weak regime. It is tested on  bootstrap data up to the octic anharmonic oscillator and works well. Here we choose the numerical bootstrap method because of its non-perturbative nature: the only artificial factor is to increase the depth of the bootstrap matrix. 

The proposed formula here for $\Delta E'(g)$~\eqref{eq:formulaP} happens to be \emph{the same formula} that we~\cite{Fan:2023bld} propose for the ground state level splitting $\Delta E_D(g)$ of double-well potentials, which captures the effect of anharmonicity in double-well potentials. In fact, we can view them as the same model $V(x)=g x^2 + x^{2n}$, with $g>0$ being oscillators and $g<0$ being double-wells respectively. Viewing $g>0$ as the phase of oscillators and $g<0$ as the phase of instantons,   the plot of  $\Delta E'(g)$ and  $\Delta E_D(g)$ in Figure~\ref{fig:all}, resembles $\lambda\operatorname{-shape}$ curves common of phase transitions. This  suggests \emph{a deep connection} between the physics of anharmonicity in the anharmonic oscillators and in the double-well potentials, from the view point of quantum phase transitions.

The oscillator physics dominates $\Delta E'(g)$ and the instanton physics dominates   $\Delta E_D(g)$, so it is \emph{unexpected} that they share the same formula~\eqref{eq:formulaP} across all coupling values. This suggests there is an analytic connection between anharmonic oscillators and double-well potentials. This suggestion is not alone. In~\cite{Caswell:1979qh}, the perturbation series of anharmonic oscillators and double-well potentials are connected by introducing an effective mass and reexpansion of series. 

We hope that our finding can serve as a guide for searching an analytic method \emph{that works both at the 'weak' and the 'strong' regime}. There are works on this direction. For example, in~\cite{IAIvanov_1998.33,IAIvanov_1998.26} energy levels in the weak and the strong regime are linked by a Laplace integral representation. The results of this paper has the potential of application to spectroscopy  and might also play a role in energy gap law in particle transfer dynamics, for example, a partial list can be found in~\cite{NitzanBook,toutounji2014new,ando1997quantum,toutounji2015electronic}.

\begin{acknowledgments}
    Wei Fan is supported in part  by the National Natural Science Foundation of China  (Grant No.12105121).  
\end{acknowledgments}
\bibliography{boot-oscillator}%
\bibliographystyle{apsrev4-1}

\end{document}